# Observation of spin-electric transitions in a molecular exchange qubit

Florian le Mardelé,[a] Ivan Mohelský,[a] Jan Wyzula,[a,b] Milan Orlita,[a] Philippe Turek,[c] Filippo Troiani*,[d] Athanassios K. Boudalis*[c]

[a] *Laboratoire National des Champs Magnétiques Intenses, CNRS-UGA-UPS-INSA-EMFL, 25 rue des Martyrs, 38042 Grenoble, France.*
[b] *Scientific Computing, Theory and Data Division, Paul Scherrer Institute, 5232 Villigen PSI, Switzerland*
[c] *Institut de Chimie de Strasbourg (UMR 7177, CNRS-Unistra), Université de Strasbourg, 4 rue Blaise Pascal, CS 90032, F-67081 Strasbourg, France. Email:* bountalis@unistra.fr
[d] *Centro S3, CNR-Istituto di Nanoscienze, I-41125 Modena, Italy. Email:* filippo.troiani@nano.cnr.it

## Abstract

Electric fields represent an ideal means for controlling spins at the nanoscale and, more specifically, for manipulating protected degrees of freedom in multispin systems. Here we perform low-temperature magnetic far-IR spectroscopy on a molecular spin triangle (**Fe$_3$**) and provide the first experimental evidence of spin-electric transitions in polynuclear complexes. The co-presence of electric- and magnetic-dipole transitions, allows us to estimate the spin-electric coupling. Based on spin Hamiltonian simulations of the spectra, we identify the observed transitions and introduce the concept of a generalized exchange qubit. This applies to a wide class of molecular spin triangles, and includes the scalar chirality and the partial spin sum qubits as special cases.

## Introduction

Manipulation of the electron spin magnetization is traditionally achieved with resonant spectroscopy techniques, such as Electron Paramagnetic Resonance (EPR). These drive the system *via* the magnetic field component of the radiation. However, multispin systems offer the possibility of going beyond this paradigm, of encoding qubits in decoherence-free subspaces, and of performing the manipulation by electrical means.[1–3]

Polynuclear magnetic molecules represent a wide class of highly engineerable multispin systems. Spin triangles – among the first and best explored polynuclear magnetic molecules[4] – have known a renewed interest in the last years, due to their possible use for the implementation of the spin chirality qubit.[5–7] Such encoding would allow the qubit manipulation by localized and rapidly oscillating electric fields, such as those generated by resonators or STM tips.[5,6] Besides, states with opposite chiralities share the same spin projections, a property that makes the chirality qubit substantially immune from hyperfine-induced decoherence.[7]

In spite of such interest, the actual observation of spin-electric transitions in polyatomic magnetic molecules has so far remained elusive. The challenge is two-fold and includes, at a general level, the identification of molecules with a sizable magneto-electric (ME) coupling and, more specifically, a fortunate correspondence between the energies of the electrically addressable transitions within the molecule and the frequency range of the experimental technique. Techniques utilizing resonators (such as EPR) are constrained to fixed frequencies and cannot be tuned at will to the excitation energies of different molecules. Moreover, they are optimized to expose the sample to only the $\mathbf{B}_1$ component of the incipient radiation. On the other hand, broadband techniques that are based on coplanar waveguides combined with microwave signal generators, are constrained to frequencies below ~100 GHz (~3 cm$^{-1}$), and are capable of addressing only a few examples of spin triangles, with very weak interactions, which are therefore unlikely to generate strong ME couplings. Similarly, these techniques are also optimized to expose the samples to the $\mathbf{B}_1$ field. Prior knowledge of the magnetism of oxo-bridged Cr$^{III}$, Fe$^{III}$ and Cu$^{II}$ spin triangles reveals that their doublet-doublet gaps are typically in the $10^1$-$10^2$ range, which falls in the far-IR (FIR) regime. Quite fortuitously, FIR spectroscopy is a quasi-optical technique, with a freely propagating light beam in which $\mathbf{E}_1$ and $\mathbf{B}_1$ are not spatially separated. This was therefore considered as the ideal means not only to access the required frequencies, but also to expose spin triangles to oscillating electric fields.

The molecule selected for this demonstration is one for which ME coupling has been experimentally observed through EPR spectroscopy. In particular, it was shown by two of the authors (A. B. and P. T.) that static electric fields of the order of $10^7$ V/m modify the intensity of the continuous-wave EPR spectra of the [Fe$_3$O(O$_2$CPh)$_6$(py)$_3$]ClO$_4$·py (**Fe$_3$**) complex.[8] Using pulsed EPR spectroscopy,[9] we have shown that pulsed electric fields can dynamically drive the magnetizations, a result that represents a significant step towards the electrical control of the spin. Subsequent theoretical[10,11] and experimental[12] investigations have confirmed the presence of a sizable ME coupling in the **Fe$_3$** molecule, while negative results have been reported for the analogous Cr$^{III}$-based spin triangle.[9] Recently, similar studies have demonstrated ME couplings in spin triangles, rings and chains.[13–15]

The **Fe$_3$** molecule also allows one to address the second challenge mentioned above. In this respect, one needs molecules whose relevant transitions (for example, the doublet-doublet energy gaps) fall in an experimentally accessible spectral region. According to our previous investigations (SQUID magnetometry and CW EPR spectroscopy)[16] the doublet-doublet gap of **Fe$_3$** should approximately be 50-55 cm$^{-1}$, i.e. within the "THz gap". This region is beyond the high-frequency EPR spectrometers[17] but are accessible to magneto-FIR (MFIR), which additionally offers very high frequency resolution. Indeed, MFIR has been successfully used to investigate mononuclear transition metal, lanthanide and actinide complexes with high zero-field splittings[15–20].

Here, we use MFIR spectroscopy to obtain the first observation of spin-electric transitions in polynuclear molecular complexes, and provide an estimate of the ME coupling. Moreover, we show that the **Fe$_3$** complex implements a generalized qubit, amenable to electrical control and characterized by a hybrid degree of freedom. This results from the simultaneous presence of magnetic frustration, Dzyaloshinskii–Moriya interactions (DMI) and unequal Heisenberg couplings, in a molecule formed by spins larger than 1/2.

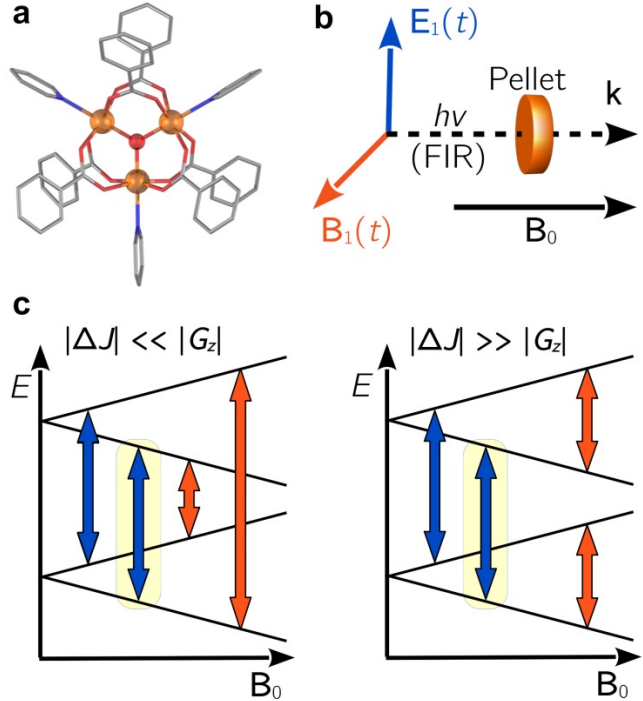

Figure 1. (a) Schematic structure of the **Fe₃** spin triangle. The orange spheres indicate Fe$^{III}$ ions and the red sphere indicates the central oxide bridge that constitutes the main superexchange pathway. Benzoato and pyridine ligands are plotted as sticks for clarity. (b) Sketch of the experimental setup, showing the far-IR beam's Poynting vector (**k**), with its electric and magnetic components ($\mathbf{E}_1(t)$ and $\mathbf{B}_1(t)$, respectively) with respect to the sample pellet and the magnetic field (Faraday geometry, $\mathbf{B}_0 \| \mathbf{k}$). (c) Schematic view of the transitions, within the ground-state quadruplet of the spin triangle, for a magnetic field $\mathbf{B}_0$ oriented along the main symmetry axis of the molecule. Orange and blue arrows correspond to magnetic- and electric-dipole transitions, respectively. Highlighted in yellow is the transition between the ground ($|0\rangle$) and the second excited state ($|1\rangle$) states that define the exchange qubit.

## From the chirality to the generalized exchange qubit

Spin triangles with dominant antiferromagnetic interaction are prototypical examples of magnetically frustrated systems. Previous investigations[5–7,18–21] have mainly focused on polyoxometalate molecules containing Cu$^{II}$$_3$[22–24] or V$^{IV}$$_3$[25–27] cores. These systems are characterized by the presence of $s_i = 1/2$ spins and by weak Heisenberg couplings and Dzyaloshinskii–Moriya interactions (DMI), mediated by long polyoxometalate "ligands". Consistently with the nominal symmetry of the molecules, the zero-field splitting between the ground and the first excited doublets is attributed to the DMI. The **Fe₃** molecule we investigate here belongs to a different family of magnetic molecules, which includes Fe$^{III}$, Cr$^{III}$ and Cu$^{II}$ triangles with monatomic central bridges[4]. At the magnetic level, **Fe₃** differs from the above mentioned polyoxometalates in two relevant respects: the length of the spins ($s_i = 5/2$) and the strength of the exchange interactions. The former aspect qualitatively affects the character of the relevant eigenstates, as detailed below. As to the latter aspect, it implies large values of the zero-field splitting, resulting from the DMI interaction and from possible inhomogeneities between the Heisenberg couplings.

At a magnetic level, all the above spin triangles can be modeled by a spin Hamiltonian that, besides the Zeeman term, also includes the Heisenberg and the Dzyaloshinskii–Moriya interactions:

$$H = (J/2)\,\mathbf{S}\cdot\mathbf{S} + (\Delta J/2)\,\mathbf{S}_{23}\cdot\mathbf{S}_{23} + G_z K_z + \mu_B \mathbf{B}_0 \cdot \tilde{\mathbf{g}} \cdot \mathbf{S} \quad (1)$$

where $\mathbf{S}_{23} = \mathbf{s}_2 + \mathbf{s}_3$ and $\mathbf{S} = \mathbf{S}_{23} + \mathbf{s}_1$ represent a partial and the total spin sums, respectively. The Heisenberg couplings between $\mathbf{s}_1$ and $\mathbf{s}_{k=2,3}$, is $J$, that between $\mathbf{s}_2$ and $\mathbf{s}_3$ is $J + \Delta J$; the DMI is expressed in terms of the $z$-component of the vector chirality, which is given by:

$$\mathbf{K}_z = (\mathbf{s}_1 \times \mathbf{s}_2)_z + (\mathbf{s}_2 \times \mathbf{s}_3)_z + (\mathbf{s}_3 \times \mathbf{s}_1)_z \quad (2)$$

The qualitative picture that results from the above spin Hamiltonian can be summarized as follows. In the presence of identical and symmetric exchange interactions ($\Delta J = G_z = 0$), the magnetic frustration results in the formation of two degenerate ground state doublets. Such a degeneracy can be partially removed by the DMI ($G_z \neq 0$). In molecules with at least $C_3$ symmetry ($\Delta J = 0$) formed by $s_i = 1/2$ spins, for a magnetic field $\mathbf{B}_0$ oriented along the main symmetry axis of the molecule ($z$), the eigenstates are characterized by well-defined values of the total spin ($S$), of its projection along the quantization axis ($S_z$), and of the vector chirality ($K_z$). An equivalent representation can be developed by replacing $K_z$ by the scalar chirality $C$, given by:

$$C = \mathbf{s}_1 \cdot (\mathbf{s}_2 \times \mathbf{s}_3) \quad (3)$$

which also commutes with the Hamiltonian and thus represents a good quantum number. In fact, $C$ and $S_z$ can be regarded as actual degrees of freedom (pseudospins) and manipulated independently within the ground quadruplet: magnetic-dipole transitions couple states of opposite spin projections ($S_z = \pm 1/2$) and identical scalar chirality, whereas electric-dipole transitions take place between states with opposite scalar chiralities ($C = \pm 1$) and identical spin projection.[5]

As detailed in the following, this picture is significantly modified for spins $s_i > 1/2$ and inhomogeneous exchange interactions ($\Delta J \neq 0$). As a result, neither the total spin, nor the scalar and vector chiralities are good quantum numbers. Nonetheless, one can manipulate the quantum state electrically within a two-dimensional sector of given spin projection (Fig. 1). In the following, we refer to such an electrically controllable degree of freedom as a *generalized exchange qubit*, which reduces to the partial-spin sum ($S_{23} = s_i \pm 1/2$) and to the scalar chirality in special cases. Interestingly, the magnetic dipole transitions within the low-energy subspace reflect the character of the exchange qubit: in fact, they occur within the two lowest doublets in the limit $|\Delta J| \gg |G_z|$, and between states belonging to different doublets in the opposite limit; both inter- and intra-doublet transitions are allowed if the two couplings are comparable in magnitude. Besides, for $\mathbf{B}_0 \| z$, independently on the ratio between $\Delta J$ and $G_z$, oscillating electric and magnetic fields induce distinct transitions.

### MFIR spectroscopy and empirical fits

Variable-field MFIR spectra carried out at 4.2 K reveal excitations with distinctively different $B$-dependencies. In particular, we observe Zeeman-like transitions, whose energies depend linearly on the field, and quasi-constant transitions, whose energies are weakly affected by $\mathbf{B}_0$ throughout the considered range of values (0-16 T). The peaks in the 20-40 cm$^{-1}$ range, which we attribute to transitions between states in the ground quadruplet, undergo a progressive splitting and shifting with increasing magnetic field (Figure 2, top left). This effect is particularly visible upon normalization to the zero-field spectrum ($T_B/T_0$ ratio, Figure 2, top right). The normalized spectra also reveal clear features around 73 and 90 cm$^{-1}$ (and at higher energies), which we tentatively assign to transitions from the ground state to states that belong to excited multiplets. Other inter-multiplet transitions display a linear (increasing or decreasing) dependence on the field, in the regions 80-100 cm$^{-1}$ and 50-65 cm$^{-1}$.

The overall dependence of the spectra on the magnetic field can be appreciated in the $v$ vs $B$ surface plot of the ratio spectra (Figure 2, bottom), which have undergone differential averaging,[28] whereby each ratio spectrum is divided by the average of the spectra within a certain magnetic field window around that specific field. This process enhances the visibility of individual field-dependent features.

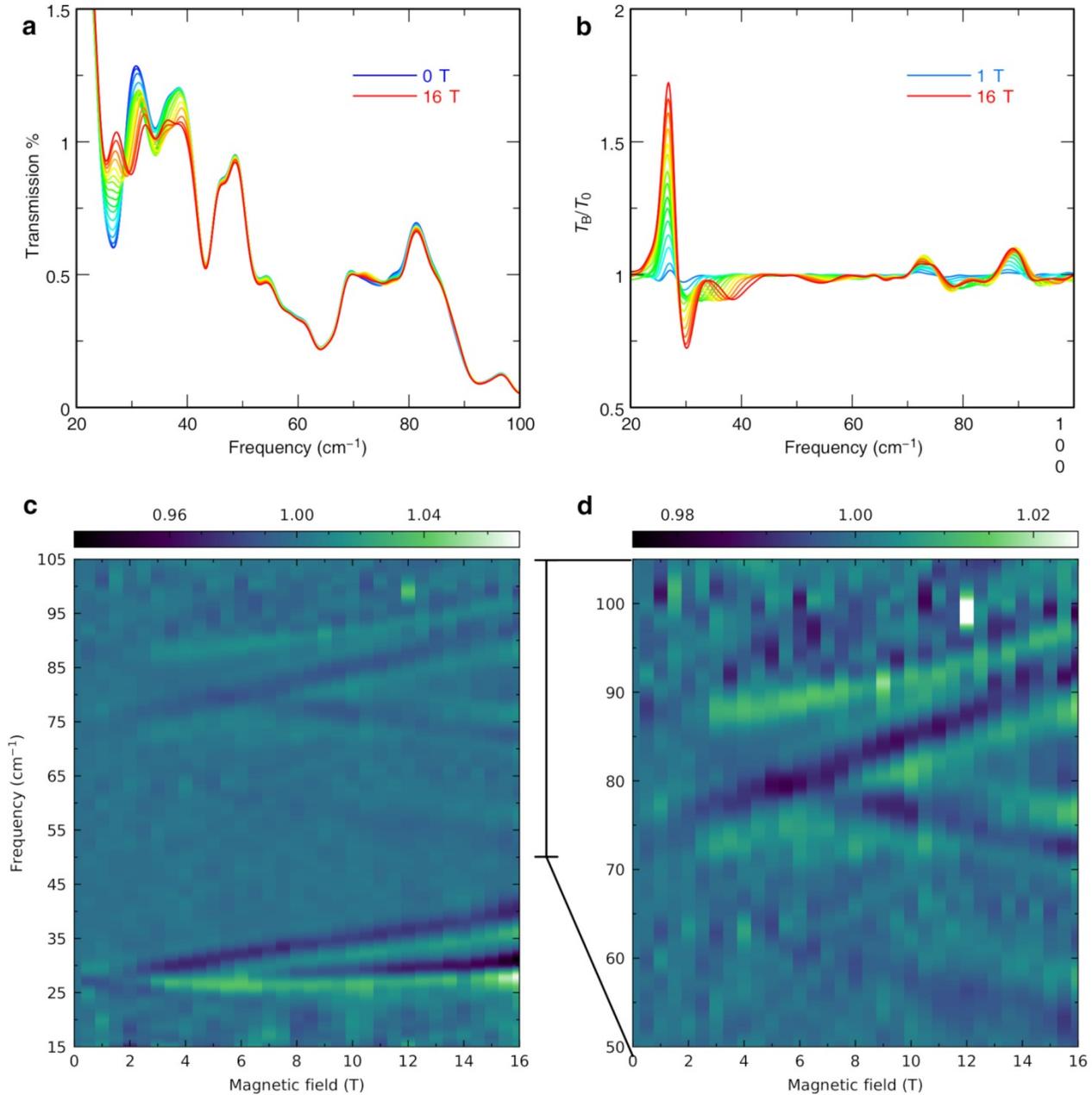

Figure 2. **(a)** FIR transmittance ($T$) spectra of **Fe₃** for various magnetic fields (**B**$_0$) between 0-16 T ($T$ = 4.2 K). **(b)** The same spectra ($T_B$), normalized by dividing with the zero-field spectrum ($T_0$). For clarity, not all intermediate fields are shown. **(c)** Surface plot of the ratio ($T_B/T_0$) MFIR spectra after differential averaging as described in the text. The low-energy (25-40 cm$^{-1}$) features are assigned to excitations to the excited doublet. **(d)** Expansion of the 50-105 cm$^{-1}$ region, assigned to excitations to higher states. The color scale is different than in (c) to accentuate the lower-intensity features.

The spectra present a joint contribution from the magnetic and the vibrational degrees of freedom. In order to discriminate between the two on a purely empirical basis, the raw spectra were fitted with a set of two field-dependent and seven field-independent Gaussian peaks. The former two peaks, characterized by a weak and a linear dependence on the magnetic field, are related to the spins, while the latter ones account for vibrations. In order to justify this assignment, and unequivocally exclude any magnetic dependence in the vibrational contributions, we carried out comparisons with the diamagnetic analogue of **Fe$_3$**, the isostructural complex [Ga$_3$O(O$_2$CPh)$_6$(py)$_3$]ClO$_4$·py (**Ga$_3$**), reported here for the first time (see SI and Figure S3). Its MFIR spectra, collected between 0-16 T, are field-independent at the level of our experimental sensitivity, definitively demonstrating that any field-dependent feature in the **Fe$_3$** spectra is of purely spin origin (see Figures S4-S6). The possible role of vibronic couplings,[29] i.e. of a coherent mixing between spin and vibrational degrees of freedom, is briefly discussed in the SI.

The above fits highlight two things: (a) they allow the isolation of the purely vibrational part of the spectrum and (b) they reveal the ratio of intensities of the magnetic and electric excitations, which is subsequently used to estimate the spin-electric coupling constant (see below).

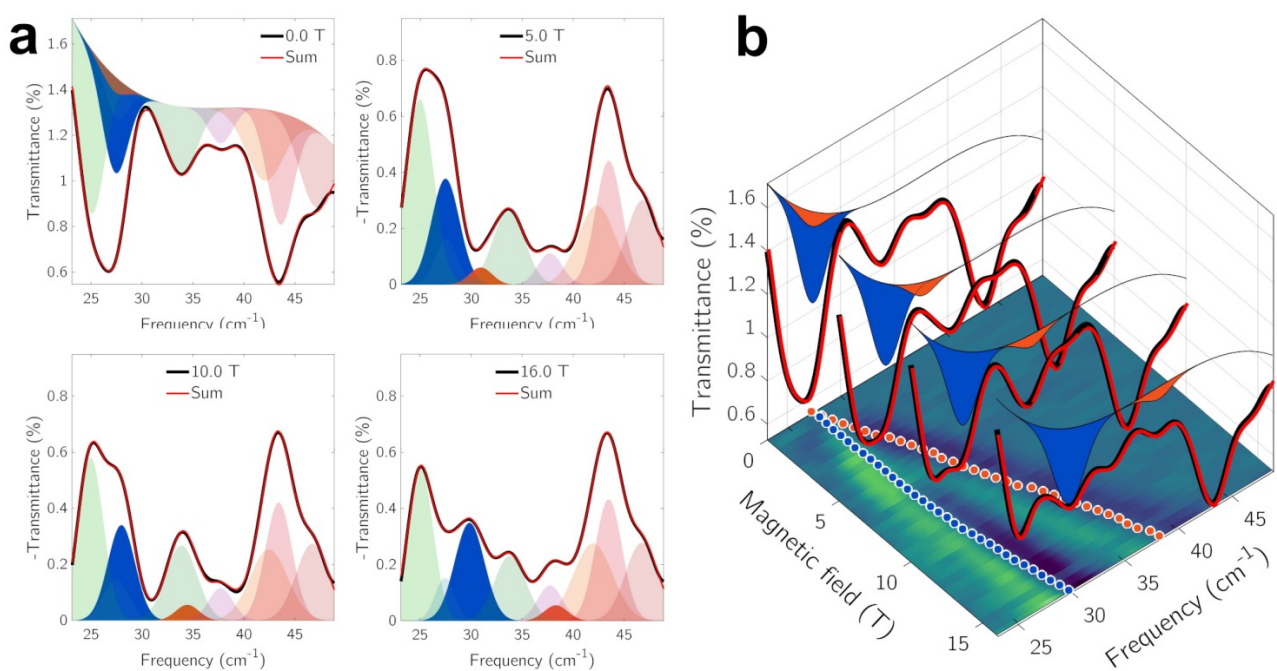

Figure 3. (**a**) Indicative fitting results to multiple Gaussians. The data manipulation is described in the legend of Figure S2. The semitransparent Gaussians are fixed-frequency absorptions assigned to purely vibrational degrees of freedom. The blue and orange Gaussians are field-shifting corresponding, respectively, to the electric- and magnetic-dipole (EPR) transitions. (**b**) 3D plot comparing the fitted Gausians to the FIR spectra at selected fields. The best-fit frequencies of the two absorptions are overlaid to the surface plot, highlighting the close agreement to the field-dependent features. The color code of the blue (electric) and orange (magnetic) transitions is the same.

## Simulated spectra and properties of the exchange qubit

In order to gain a quantitative understanding of the observed features, we simulated the MFIR spectra as a function of the magnetic field. The molecules in the sample present a random orientation with respect to the static magnetic field **B**$_0$, and to the oscillating electric (**E**$_1$) and magnetic (**B**$_1$) fields that induce the observed transitions. The tilting angle $\theta$ between the magnetic field and the main symmetry axis of the molecule affects both the energy and the amplitude of the transitions, resulting respectively from the eigenvalues and eigenstates of the spin Hamiltonian (Eq. 1). It should also be noted that the rotations around the molecular $z$-axis affect the spin-electric couplings and the related transitions probabilities.

As a peculiar feature in the present investigation, observed spectra result both from electric-dipole and from magnetic-dipole induced transitions. These provide contributions that are comparable in magnitude, though with qualitatively different features. In the simulations, the magnetic- and electric-dipole transitions are induced by the Hamiltonians:

$$H_B = g\mu_B \mathbf{B}_1 \cdot \mathbf{S} \quad (4)$$

and

$$H_E = \kappa \sum_{i=1}^{3} \left( \mathbf{E}_1 \cdot \hat{\mathbf{u}}_k \right) \mathbf{s}_i \cdot \mathbf{s}_j \quad (5)$$

respectively (where $i$ and $j$ differ from $k$ and from each other). Here, $\kappa$ represents the spin-electric coupling constant and the $\mathbf{u}_k$ are the unit vectors perpendicular to the side of the triangle that connects the spins $i$ and $j$.

As detailed in the Methods and in the SI, the simulated spectra are obtained by considering an initial thermal state with $T = 4.2$ K and by performing a spherical average of the spectra to account for the random orientation of the molecules within the powder sample. In order to determine a suitable set of Hamiltonian parameters, we have preliminarily compared the energy eigenvalues (referred to the ground state energy) with the main transition energies emerging from the normalized MFIR spectra ($T_B/T_0$), as shown in Figure S1. The simulation of the overall spectra subsequently allowed a fine tuning of the parameters: a very good agreement was finally obtained for $J = 43.8$ cm$^{-1}$, $\Delta J = -2.2$ cm$^{-1}$ and $G_z = -1.9$ cm$^{-1}$.

The computed energies and amplitudes of the transitions related to the magnetic- ($H_B$) and electric-dipole ($H_E$) couplings allow us to assess the occurrence of both electric- and magnetic-dipole transitions and to estimate the corresponding contributions. The presence of molecules with different orientations relative to the fields and the fact that $|\Delta J|$ is comparable to $|G_z|$ give rise to an overlap between the two kind of transitions, which are disjoint in special cases (e.g., for $\mathbf{B}_0 \| z$). However, the two contributions, hereafter referred to as $I_B$ and $I_E$, still maintain peculiar and distinguishable features, and are both required in order to reproduce the observed spectra (Figure 4). Besides, their co-presence can be exploited in order to estimate the value of the ME coupling constant $\kappa$, whose value is not experimentally accessible otherwise. In particular, the fact that $I_B$ and $I_E$ are proportional respectively to $(g\mu_B B_1)^2$ and $(\kappa E_1)^2$, combined with the calculation of the transition amplitudes and with the estimate of the electric ($E_1$) and magnetic ($B_1$) components of the oscillating field in the sample, leads to the value $\kappa \approx 3.97 \times 10^{-4}$ e nm, which is in line with the values derived for other molecular spin triangles.[19,30]

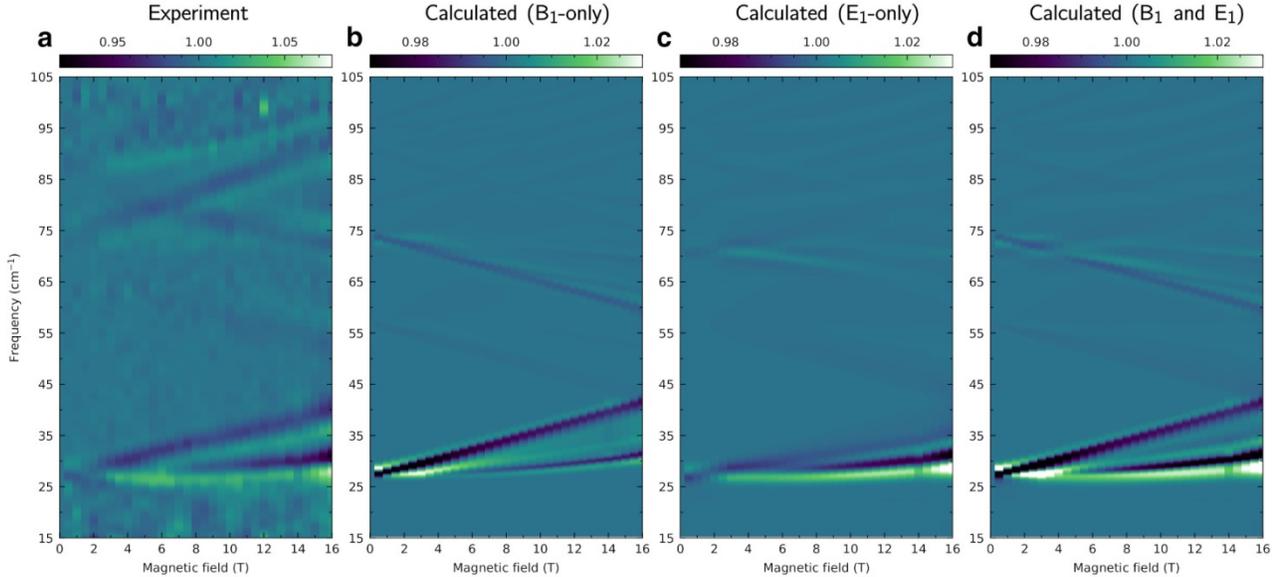

Figure 4. (**a**) Experimental surface plot of the differential averages of the ratio MFIR spectra of **Fe₃**. (**b**)-(**d**) Simulations of the powder-averaged spectra considering the static spin Hamiltonian and the absorptions induced by: (**b**) Only the oscillating field **B**$_1$. (**c**) Only the oscillating field **E**$_1$. (**d**) Both fields **B**$_1$ and **E**$_1$. Each surface plot was calculated by a differential averaging of the ratio ($T_B/T_0$) spectra. In the case of (c), the **B**$_1$- and **E**$_1$-induced components were scaled by factors $\alpha_B = 2$ and $\alpha_E = 0.45$, respectively.

Spin-electric transitions potentially represent a novel tool to investigate magnetic molecules, possibly a complementary one with respect to conventional EPR. As discussed above, the observation of these transitions does not necessarily require a specific orientation of the molecule with respect to the magnetic field, nor the presence of a $C_3$ (or higher) symmetry or of a well-defined chirality. On the other hand, some of these conditions are desirable in order to make the electric-dipole and the magnetic-dipole transitions complementary, and to dramatically improve the coherence properties of the spin qubit. Hereafter, we briefly analyze these aspects and their implications.

Irrespective of value $\Delta J$, the presence of DMI in a spin triangle formed by spins $s_i > ½$, gives rise to quantum fluctuations in total spin, in the scalar and in the vector chiralities: none of these quantities corresponds to a good quantum number, as can be formally deduced from the expressions of the commutators between the H on the one hand, and $\mathbf{S}^2$, $C$ and $K_z$ on the other (see the SI). The quantum fluctuations that characterize these operators can be numerically quantified by their standard deviations, $\Delta O \equiv [\langle O^2 \rangle - \langle O \rangle^2]^{1/2}$, which are reported in Table 1 for the ground ($|0\rangle$) and second excited ($|1\rangle$) states. As a result, unlike for triangles formed by $s_i = ½$ spins, the intra-multiplet transitions induced by the spin-electric coupling cannot be exactly regarded as spin-chiral. However, such deviation has no significant implication for the implementation of the exchange qubit.

As already shown for the case of the chirality qubit,[6] the orientation of the magnetic field along the main symmetry axis of the molecule ($z$) allows to decouple the exchange qubit from the total spin projection. For nonzero tilting angles $\theta$, it is no longer possible to identify, within the ground quartet, subspaces corresponding to identical values of the total spin projection along the main molecule or along the quantization axis ($S_\theta = S_z \cos\theta + S_x \sin\theta$). This implies not only an overlap between electric- and magnetic-dipole induced transitions, but also a coupling of the qubit with the nuclear-induced magnetic noise, which represents the main source of spin decoherence at cryogenic temperatures.

As to the presence of unequal exchange couplings ($\Delta J \neq 0$), this does not generally prevent the possibility of observing spin-electric transitions. In fact, even in the limiting case where $|\Delta J| \gg |G_z|$, the transition amplitude between the two qubit states is nonzero, and can be analytically expressed as:

$$\langle S_{23} = s_i - 1/2 | H_E | S_{23} = s_i + 1/2 \rangle = \frac{\sqrt{3}}{4}(s_i + 1/2)\kappa \mathbf{E}_1 \cdot (\hat{\mathbf{u}}_2 - \hat{\mathbf{u}}_3) \quad (6)$$

where both states are characterized by $S = 1/2$ and $M = -1/2$. On the other hand, the decoupling of the exchange qubit from the nuclear spins essentially depends on the indistinguishability of the $|0\rangle$ and $|1\rangle$ states in terms of spin polarization. This implies that $\langle 0|s_i|0\rangle = \langle 1|s_i|1\rangle$ for each of the three spins (see Table 1): a condition that is approached in the limit $|G_z| \gg |\Delta J|$, and fully met only for $\Delta J = 0$.[7]

Table 1. Expectation values and standard deviations ($\Delta$) of the scalar and vector chiralities and for the total spin corresponding to the two qubit states, i.e. the ground and second excited states of the spin Hamiltonian $H$. The values refer to the case $B_0 = 1$ T, $\theta = 0$. The values in parentheses are obtained by setting $\Delta J = 0$. The effect of the signs of $\Delta J$ and $G_z$ on the expectation values and standard deviations of the scalar and vector chiralities is presented in Figures S7 and S8.

|  | $\langle C \rangle / \hbar^3$ | $\Delta C / \hbar^3$ | $\langle K_z \rangle / \hbar^2$ | $\Delta K_z / \hbar^2$ | $\langle \mathbf{S}^2 \rangle / \hbar^2$ | $\Delta \mathbf{S}^2 / \hbar^2$ | $\langle s_{1z} \rangle$ | $\langle s_{2z} \rangle = \langle s_{3z} \rangle$ |
|---|---|---|---|---|---|---|---|---|
| $|0\rangle$ | -8.66 | 2.27 | 14.1 | 9.52 | -0.204 | 0.737 | 0.0511 | -0.276 |
|  | (-8.94) | (0.246) | (14.2) | (9.40) | (0.707) | (0.656) | (-0.166) | (-0.166) |
| $|1\rangle$ | 8.60 | 2.38 | -0.346 | 12.5 | -0.0605 | 0.963 | -0.384 | -0.0578 |
|  | (8.89) | (0.335) | (-0.76) | (12.5) | (0.941) | (0.860) | (-0.166) | (-0.166) |

## Conclusions and outlook

In conclusion, our study on **Fe₃** provides the first observation of spin-electric transitions in polynuclear magnetic molecules. The co-presence of magnetic- and electric-dipole induced transitions with comparable intensities, is a feature that we exploit for estimating the magneto-electric coupling. Besides, we theoretically demonstrate that **Fe₃** can be regarded as a generalized exchange qubit, sharing characteristics of the spin-chirality and of the partial-spin sum qubits.

The use of crystal samples, where the alignment of the magnetic field with the molecule axis can be achieved for all the molecules in the ensemble, will provide an even clearer insight and allow a more selective addressing of the spin-electric transitions. This geometry would also allow the observation of extended "clock" transitions, which would be globally (i.e. not only in the vicinity of an avoided level crossing) insensitive to fluctuations of the external and nuclear-induced magnetic fields.

The observation of spin-electric transitions represents a significant step towards an all-electric manipulation of the molecular spin. Pulsed excitation experiments in the THz regime, will be required in order to measure the relaxation and coherence times of the molecular exchange qubit. Further studies should also elucidate the matrix effects in crystalline samples and in molecular layers deposited on surfaces.

## Data availability

Crystallographic data for **Ga₃**, have been deposited in cif format with the CCDC under deposition number 2334338. MFIR data are available by the authors upon reasonable request.